\documentclass[12pt,preprint]{aastex}
\input epsf.tex
\begin{document}
\title{THE WHITE DWARF COOLING SEQUENCE OF THE GLOBULAR CLUSTER MESSIER 4\altaffilmark{1}}
\author{Brad M. S. Hansen\altaffilmark{2,3}, James Brewer\altaffilmark{4}, Greg G. Fahlman\altaffilmark{4,5}, Brad K. Gibson\altaffilmark{6},
Rodrigo Ibata\altaffilmark{7}, Marco Limongi\altaffilmark{8}, R. Michael Rich\altaffilmark{2}, Harvey B. Richer\altaffilmark{4},
Michael M. Shara\altaffilmark{9}, Peter B. Stetson\altaffilmark{10}
}
\altaffiltext{1} {Based on observations with the NASA/ESA Hubble Space Telescope, obtained at the Space Telescope Science
Institute, which is operated by the Association of Universities for Research in Astronomy, Inc., under NASA contract NAS 5-26555.
These observations are associated with proposal GO-8679}
\altaffiltext{2}{Hubble Fellow,  Department of Astronomy, University of California Los Angeles, Los Angeles, CA 90095, hansen@astro.ucla.edu, rmr@astro.ucla.edu}
\altaffiltext{3}{Department of Astrophysical Sciences, Princeton University, Princeton, NJ 08544}
\altaffiltext{4}{ Department of Physics \& Astronomy, 6224 Agricultural Road, University of British Columbia, Vancouver, BC, V6T 1Z4, Canada, richer@astro.ubc.ca }
\altaffiltext{5}{ Canada-France-Hawaii Telescope Corporation, P.O. Box 1597, Kamuela, HA, 96743, fahlman@cfht.hawaii.edu}
\altaffiltext{6}{ Centre for Astrophysics Supercomputing, Mail No. 31, Swinburne University, P.O.Box 218, Hawthorn, Victoria 3122, Australia, bgibson@astro.swin.edu.au}
\altaffiltext{7}{ Observatoire de Strasbourg, 11, rue de l'Universite, F-67000 Strasbourg, France, ibata@newb6.u-strasbg.fr}
\altaffiltext{8}{ Osservatorio Astronomico di Roma, Via Frascati 33, Monteporzio Catone I-00040, Italy, marco@nemo.mporzio.astro.it}
\altaffiltext{9}{ Department of Astrophysics, Division of Physical Sciences, American Museum of Natural History, Central Park West at
79th St, New York, NY, 10024-5192, mshara@amnh.org}
\altaffiltext{10}{ National Research Council, Hertzberg Institute of Astrophysics, 
5071 West Saanich Road, RR5, Victoria, BC, V9E 2E7, Canada, peter.stetson@nrc.ca}

\lefthead{Hansen {\it et al.}}
\righthead{M4 White Dwarf Sequence}

\begin{abstract}
We present the white dwarf sequence of the globular cluster M4, based on a 123~orbit Hubble Space Telescope exposure, with
limiting magnitude V$\sim 30$, I$\sim 28$. The white dwarf luminosity function rises sharply for I$>25.5$, consistent with the behaviour expected
for a burst population. The white dwarfs of M4 extend to approximately 2.5 magnitudes fainter than the peak of the local
Galactic disk white dwarf luminosity function. This demonstrates a clear and significant age difference between the Galactic
disk and the halo globular cluster M4. Using the same standard white dwarf models (Hansen 1999) to fit each luminosity function yields ages of
 $7.3 \pm 1.5$~Gyr for the disk and $12.7 \pm 0.7$~Gyr for M4 (2$\sigma$ statistical errors).
\end{abstract}


\section{Introduction}

Globular clusters have been the fundamental laboratories for tests of stellar structure theories 
ever since the work of Sandage (1953). A by-product of such studies is an age determination
for the cluster which is of cosmogonical interest by virtue of the large value. The study of the main
sequence and giant branches has become a detailed and quantitative field (Stetson, Vandenberg \& Bolte 1996;
Sarajedini, Chaboyer \& Demarque 1997) and the
ages thus determined offer a lower limit to the age of the Universe.
In contrast, the study of the white dwarfs in globular clusters has progressed much more slowly. 
Their low luminosities hampered early searches (Richer 1978; Ortolani \& Rosino 1986; Richer
\& Fahlman 1988; Paresce, De Marschi \& Ramoniello 1995; Elson {\it et al.} 1995).
Richer {\it et al.} (1995; 1997) reported the first detection of a significant globular cluster white dwarf population
in the cluster M4. Here we report the first results of a very deep exposure of the outer field of that
study, allowing
both the detection of very faint white dwarfs and the removal of background stars and galaxies by virtue of the
cluster proper motion.

The motivation for this work is severalfold. The contemporaneous and chemically homogeneous nature of
the cluster sample makes it an excellent laboratory to study the physics of white dwarf cooling,
just as it has been for main sequence and giant stars. The determination of an age from the white dwarf
luminosity function is particularly attractive because it allows a direct comparison, using the
same method, between the cluster and the Galactic disk. Furthermore, the internal physics of white dwarfs
is considerably different from that of main sequence stars. In particular, the sensitivity of the
atmospheric models to metallicity is expected to be unimportant as all elements heavier than helium
sink under the influence of the high white dwarf gravity. Thus, a comparison of ages determined
from white dwarf and main
sequence models offers an excellent test of systematic model uncertainties.

Finally, the continuing interest in the existence of a halo
white dwarf population makes the cluster white dwarfs useful as templates to guide searches for
similarly old white dwarfs which could potentially contribute to the Galactic dark matter.

\section{The Cluster White Dwarf Sequence}
\label{Observations}

The data are drawn from Hubble Space Telescope exposures of M4 in programs GO~5461 (cycle~4)
and GO~8679 (cycle~9) and are described in detail in the preceding paper, Richer {\it et al.} (2002).
The photometric solutions are performed by the method of profile fitting. The deep second
epoch is used to fix the position centroid for cluster members. The position
(in the cluster frame of reference) is
then held fixed when determining the magnitude in the (shallower) first epoch data. This
allows the detection and measurement of much fainter objects than in Richer et al. (1997). The final
sample consists of objects that have been detected in at least three data sets (where
the four possible sets are the V and I data at each of the two epochs). Completeness
corrections were determined using artificial star tests and require that the added
stars be recovered within 0.5~pixels of their original positions and within 0.5 magnitudes
of their original input magnitude. 

The entire cluster white dwarf cooling sequence is shown in Figure~1.
The resulting V-band luminosity function is shown in Figure~2. Also
shown is the disk white dwarf luminosity function from Liebert, Dahn \&
Monet (1988), using the improved photometry of Leggett, Ruiz \& Bergeron (1998)
and a V-band distance modulus of 12.51. This diagram provides the clearest and
most direct evidence of a significant age difference between the Galactic disk and
the Galactic halo cluster M4, predicated only on the assumption that white dwarfs cool
as they age. While this may not surprise many astronomers, until now such
evidence has been remarkably unconvincing (e.g. Carraro, Girardi \& Chiosi 1999).
 While (halo) globular cluster ages have
been estimated at $>10$~Gyr (Chaboyer {\it et al.} 1998; Cassisi  {\it et al.}  1999; Caretta  {\it et al.}  2000; Thompson  {\it et al.}  2001) for some time, age estimates (ignoring, for the moment,
white dwarf-based estimates, to which we shall return) for the local
Galactic thin disk range from 8-15~Gyr (Ng \& Bertelli 1998; Jimenez, Flynn \& Kotonova 1998;
Liu \& Chaboyer 2000; Binney, Dehnen \& Bertelli 2000).

Another striking aspect of the M4 luminosity function is the sharp rise
in the number counts from $V \sim 26.5$ to $V \sim 27.5$. This is the expected
characteristic behaviour of the white dwarf population resulting from a
single burst of star formation (e.g. Hansen 2001) and is to be contrasted
with the more gradual rise and turnover at the faint end expected from a
more extended period of star formation.

\section{ Comparison with the models}
\label{Models}

To obtain a quantitative measure of this age difference, we must turn to 
theoretical models of white dwarf cooling. A complete parameter study is
beyond the scope of a {\it Letter}, but we may obtain a first estimate by using
the current default set of models. It is also important that we redo
the age estimate for the local white dwarfs with the same models
and in the same way in order
to be consistent.

We shall assume models with carbon/oxygen cores, using the mixtures
of Segretain \& Chabrier (1993). We shall assume a chemical composition 
uniform in layers (`onion-skin model') due to gravitational sedimentation,
with helium surface layers of mass $M_{\rm He} \sim 10^{-2} M_{\rm wd}$. We will
consider models both with no surface hydrogen layers and with a 
hydrogen mass fraction of $10^{-4}$ on the surface.
 The main sequence ages are taken from the models
of Hurley, Pols \& Tout (2000) and the white dwarf models from Hansen (1999).
The main sequence-white dwarf mass relation is that of Wood (1992).

The significant difference in the calculation of the cluster and disk
luminosity functions is the choice of initial mass function and star formation
history. 
The star formation rate in the Galactic disk is assumed
to be constant and the IMF is assumed to be Salpeter (slope $\alpha$=2.35). For M4, we have assumed
the star formation is a single burst and the mass function (in the white dwarf
progenitor range) slope is $\alpha$=1.05,
based on comparing the white dwarf number count with the main sequence counts in our
field (see Richer  {\it et al.}  2002). The derived age is not sensitive to the value of
$\alpha$ in the range 0.7 to 1.1. 

The other potentially significant uncertainty is the fraction ($f_H$) of white
dwarfs that cool with hydrogen atmospheres versus those that cool with helium
atmospheres. For the local Galactic disk, this fraction is estimated to
be $f_H \sim 0.5$--0.7 (Sion 1984; Greenstein 1986; Bergeron, Ruiz \& Leggett 1997). However, the origin of this dichotomy
is not well understood (see Bergeron, Ruiz \& Leggett 1997), so we will
fit for two parameters in our luminosity function, $f_H$ and $T_{gc}$,
the age. The result is shown in Figure~3. We find an age of
$T_{gc} = 12.7 \pm 0.7$~Gyr (2$\sigma$), or a lower limit of 12~Gyr. The age determination
is not sensitive to $f_H$ as long as $f_H>0.5$. This is easily understood
because helium atmosphere white dwarfs cool very rapidly after ages $\sim 5$~Gyr and
the  best-fit pure helium atmosphere luminosity functions are a very poor fit to the data.
As noted above, the artificial star tests used to determine the incompleteness fraction incorporated
constraints on both positional recovery and magnitude accuracy. However, the formal error
bars do not include systematic model uncertainties beyond the most important one (the  parameter $f_H$).
These will be addressed in a future publication.
 Figure~4 shows a direct comparison between
the data and three theoretical luminosity functions (multiplied by the estimated incompleteness
function) for ages of 10, 12.5 and 15~Gyr. 

The error budget in most globular cluster age determinations is dominated by
the distance uncertainty (e.g. Bolte \& Hogan 1995). We shall discuss the
influence of this in detail in a subsequent publication, but we note
that we have used the distance and reddening determined by Richer  {\it et al.}  (1997)
using subdwarf main sequence fitting. Preliminary investigations show that 
changes of 0.2~magnitudes in the distance modulus (in either direction) do not change the lower bound
of the age determination, although the value of the $\chi^2$ statistic (which
is $\sim 1$ per degree of freedom for our best fit models) does become
worse. We attribute this robustness to the fact that we are fitting the
complete cooling function, rather than trying to identify a particular
localised feature.

We have performed the same kind of analysis (allowing the fraction $f_H$ to vary) on the Galactic disk luminosity
function of Liebert, Dahn \& Monet (1988) but using the improved
photometry of Leggett, Bergeron \& Ruiz (1998). 
The resulting disk
age is $7.3 \pm 1.5$~Gyr. The low age is consistent with the fact that
the faintest luminosity bins are dominated by helium atmosphere white
dwarfs, which cool faster than hydrogen atmosphere dwarfs (for a detailed discussion see Hansen 1999).
It is also worth noting that the age will increase if the Luyten survey is
incomplete at the faint end. Recent tests indicate that this is not a
concern (Monet et al 2000).
This result suggests a significant age difference between the formation
of the stellar halo of the Galaxy and of the formation of the stars in the local Solar
neighbourhood.

We should note that two other determinations of the Galactic disk
luminosity function (Oswalt  {\it et al.}  1996; Knox  {\it et al.}  1999) produce a  more
gradual cutoff than the Liebert  {\it et al.}  sample, and consequently lead
to larger ages (8 -- 12~Gyr). However, the objects in these samples do not have published
individual
atmospheric composition determinations. The resultant uncertainty in the
bolometric corrections allows a large range of age estimates.
 Thus we favour the luminosity function of Leggett  {\it et al.}  (1998)
in our age determination.

\section{Discussion}

We find an age for the globular cluster M4 of $12.7 \pm 0.7$~Gyr using the
white dwarf cooling sequence and the models of Hansen (1999). This method is
completely independent of the main sequence turnoff age method, which yields
an average age of $13.2 \pm 1.5$~Gyr for the metal poor globular clusters
(Chaboyer 2001). Our age for M4 is also in excellent agreement with a third
very different stellar chronometer, namely the nuclear decay measurements of
uranium and thorium, which yields $12.5 \pm 3$~Gyr for a metal-poor halo
star (Cayrel  {\it et al.}  2001).

Using the same method and models on the  Liebert  {\it et al.}  (1988) and Leggett
 {\it et al.}  (1998) data for the Galactic disk yields
a local disk age $7.3 \pm 1.5$~Gyr.
This implies a delay $>3$~Gyr between the formation of the
Galactic halo and the stars of the local Solar neighbourhood. Liu \& Chaboyer (2000)
find a similar age difference between the local metal-rich Hipparcos stars and
the thick disk globular cluster 47~Tuc.

To place these results in a cosmogonical context, we note that calculations of
the self-regulated gravitational settling of an initially hot protodisk show
 delays of 4-6~Gyr before stars begin to form in a true
thin disk (Burkert, Truran \& Hensler 1992). Thus, our results find a natural
explanation in this picture, where the globular clusters accompany the formation
of the Galactic halo, infalling gas settles into a hot, extended protodisk
which produces a lower level of star formation 
 for several~Gyr
until the gas has cooled sufficiently to form a true thin disk and begin
star formation in earnest. It is the latter population which we then
identify with the local white dwarf luminosity function.

This expectation can also be translated into cosmological terms by
asking at what redshift we expect true disk galaxies to form?
We will adopt
the current best fit (flat) cosmological model in which $\Omega_{\Lambda}=0.7$,
$\rm \Omega_{m}=0.3$ and $\rm H_0 = 70 \, {\rm km s^{-1} Mpc^{-1}}$. The total age is then 14~Gyr, which is consistent
with the prompt formation of M4 at $3\sigma$. At $1\sigma$, we find a delay of at least 0.9~Gyr,
and the best fit age of 12.7~Gyr suggests formation at $z \sim 6$.
For the Galactic disk, an age $<9$~Gyr requires formation at $z<1.5$. Current observations
are approaching this limit (Vogt  {\it et al.}  1997; Van Dokkum \& Stanford 2001) and
the evidence for large velocity widths in quasar absorption line studies (Prochaska \& Wolfe 1997)
would exceed it, although there are other potential explanations in this case
(Wolfe \& Prochaska 2000). In this scenario then, the epoch ($z \sim 3$) marked by the appearance
of starbursting galaxies and galactic outflows (e.g. Shapley  {\it et al.}  2001; Frye, Broadhurst \&
Benitez 2001) is associated
with the intermediate period in which the hot protogalactic disk is cooling and forming
stars, and before true `thin disk' star formation begins (this may also be the epoch of
bulge formation).

It is interesting to speculate whether the high velocity white dwarf sample of
Oppenheimer  {\it et al.}  (2001) is the missing white dwarf population from this intermediate epoch. 
 There is considerable debate about whether it is 
indeed a halo sample as originally claimed (Oppenheimer  {\it et al.}  2001; Koopmans \& Blandford 2001)
or a thick disk tracer (Reid, Sahu \& Hawley 2001). Hansen (2001) has
also argued that the age distribution suggests a more extended star formation
history than the short burst expected from a true halo population. In the latter
case, the luminosity function should look quite similar to that of M4 (modulo
differences in the IMF). In particular, the sharp rise of the M4 data at the
same point as the disk LF starts to fall off illuminates the need to go well
beyond the limits of the disk LF before making quantitative estimates of the
halo white dwarf fraction.

\acknowledgements
RMR \& MS acknowledge support from proposal GO-8679, and BH from a Hubble
Fellowship HF-01120.01, which were provided
by NASA through a grant from the Space Telescope Science Institute, which
is operated by the
 Association of Universities
for Research in Astronomy, Inc., under NASA contract NAS5-26555.
MR also acknowledges the hospitality of the University of British Columbia,
where some of this research was done.
The research of HBR \& GGF is supported in part by 
the  Natural Sciences and Engineering Research Council of Canada.
 HBR extends his appreciation to the Killam Foundation and
the Canada Council through a Canada Council Killam Fellowship. 
BKG acknowledges the support of the Australian Research Council through its
Large Research Program A00105171.

\newpage

\figcaption[cwd.ps]{The white dwarf sequence for the globular cluster M4 shows the
sharp rise in the number counts at the faint end, characteristic of a burst
population. Dashed lines show the limit for 50\% recovery fraction in at least three out of
the four data sets (V and I, each at two epochs). Filled circles
have moved at most 0.25 pixels relative to the cluster mean motion centroid and open circles
have moved at most 0.5 pixels relative to the cluster mean motion centroid. The centre of the inner halo,
for comparison, has moved $\sim 1$~pixel with respect to the cluster centre. The red (helium atmosphere)
 and blue (hydrogen atmosphere) curves
show representative 0.6$M_{\odot}$ white dwarf cooling models.
}

\figcaption[LFI4.ps]{The filled circles are the raw M4 data (for a proper motion cut $< 0.5$~pixels relative to
the cluster centre). The open circles show the counts when  corrected for incompleteness.
The filled triangles are the ($V_{max}$-weighted) disk sample of
Liebert, Dahn \& Monet (1988), using the improved photometry of Bergeron, Ruiz \& Leggett (1998).
A V-band distance modulus of 12.51 has been used (although the points are shifted by -0.1~mag in
order to not overlap the cluster points) and an arbitrary vertical shift has been applied.
}

\figcaption[ft2.ps]{The 1, 2 and 3$\sigma$ confidence intervals for the M4 age determination
are shown by the solid contours.
The dashed contour represents the $2 \sigma$ confidence interval based on the luminosity
function of Leggett, Bergeron \& Ruiz (1998), although the
$V_{max}$ weighting procedures used in constructing the  luminosity functions means this is only a rough estimate.
 The vertical dotted line indicates a value $f_H=0.7$, which is
believed to apply to the local disk white dwarf sample. The right hand axis shows the redshift corresponding to
the lookback time in the cosmological model described in the text. The left-hand error bars show the 2$\sigma$
age ranges.
}

\figcaption[LFI3_7.ps]{The solid points are the raw M4 counts. The solid
histogram represents the theoretical luminosity function for $f_H=0.7$ and
$T_{gc}=12.5$~Gyr. Dotted and dashed histograms are the same but for ages
of 10 and 15~Gyr respectively. The last data point i.e. the shaded region,
is not used in the fit because of the large incompleteness correction factor applied (Richer  {\it et al.}  2002).
}

\clearpage
\plotone{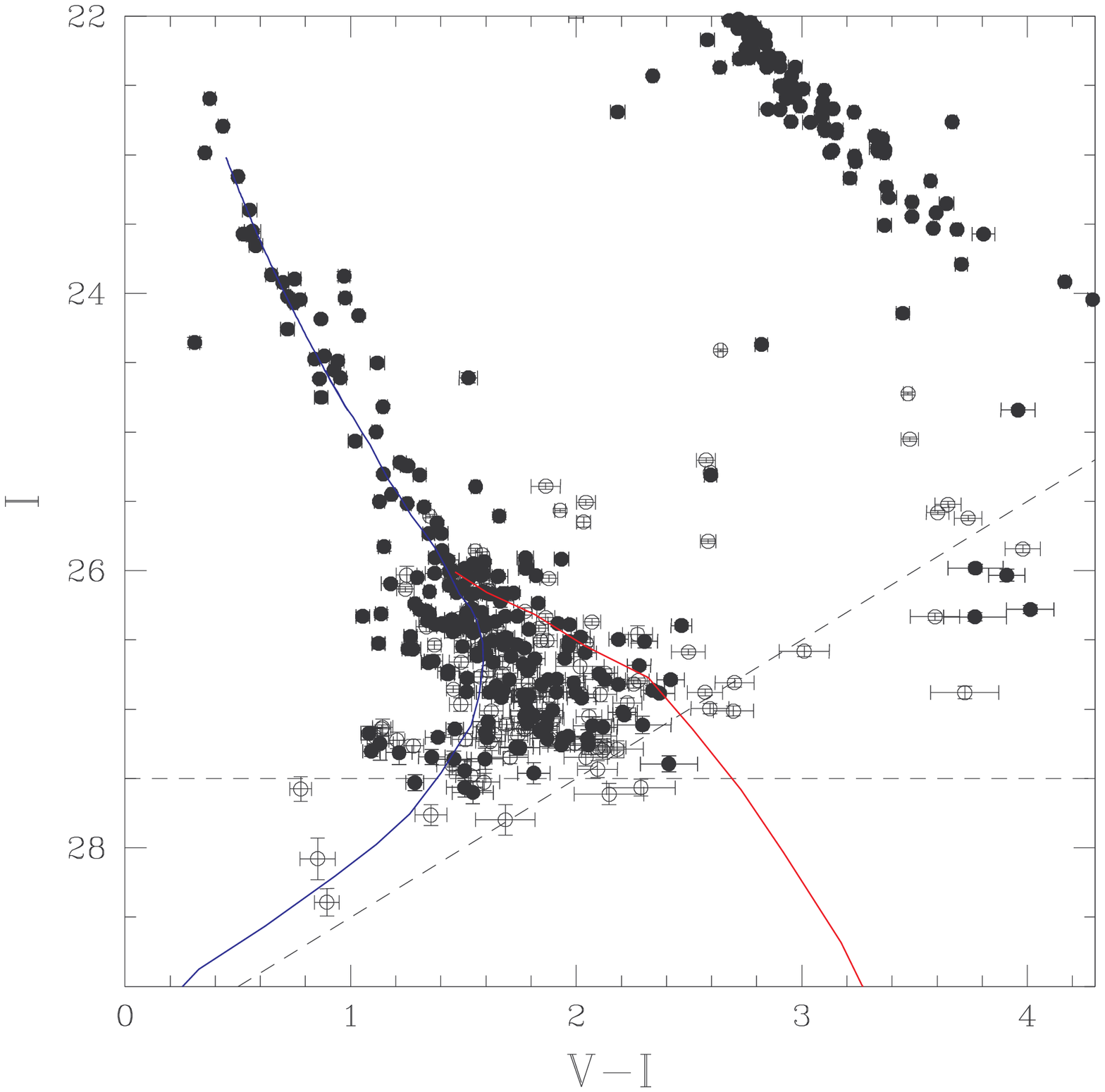}
\clearpage
\plotone{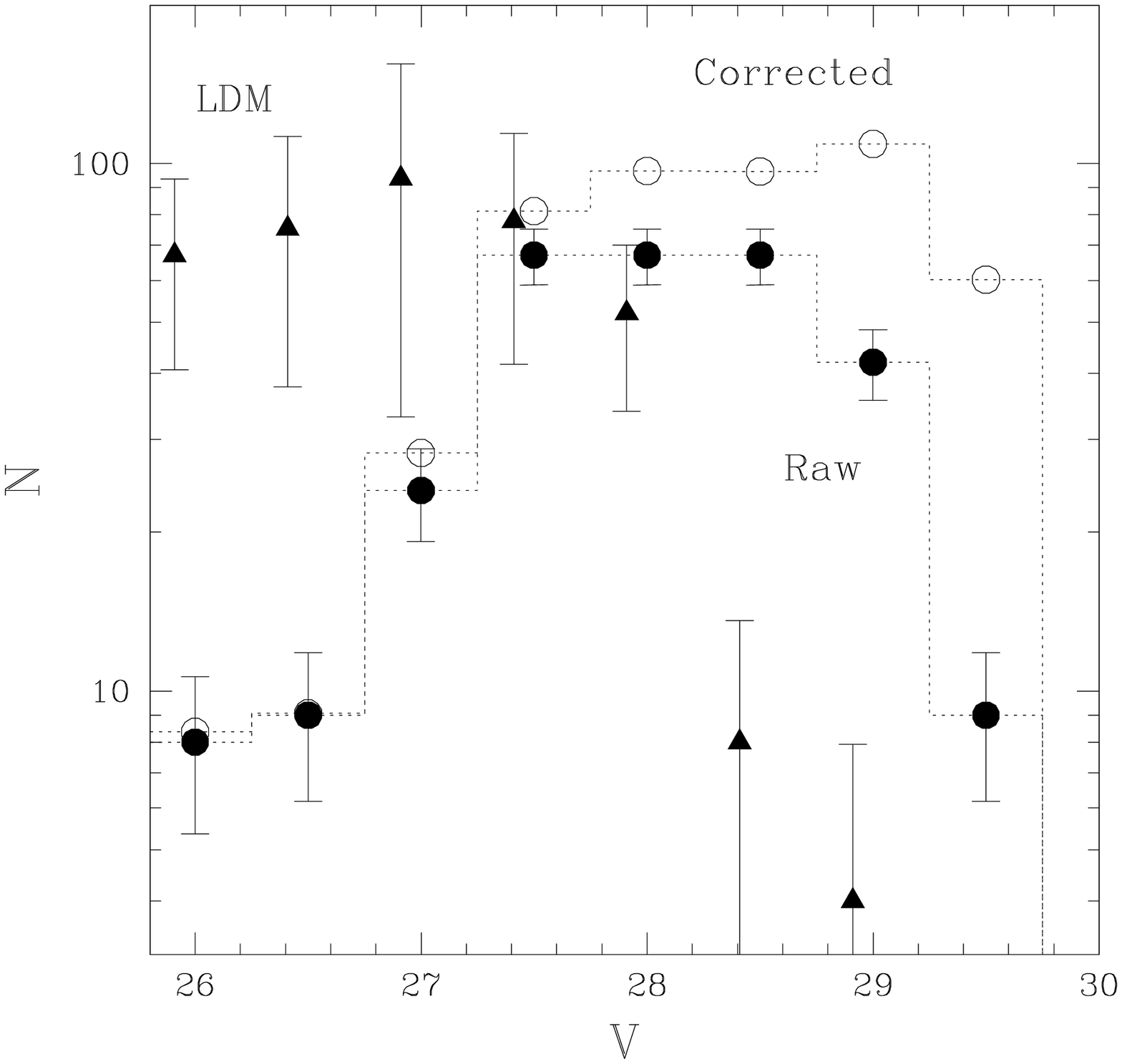}
\clearpage
\plotone{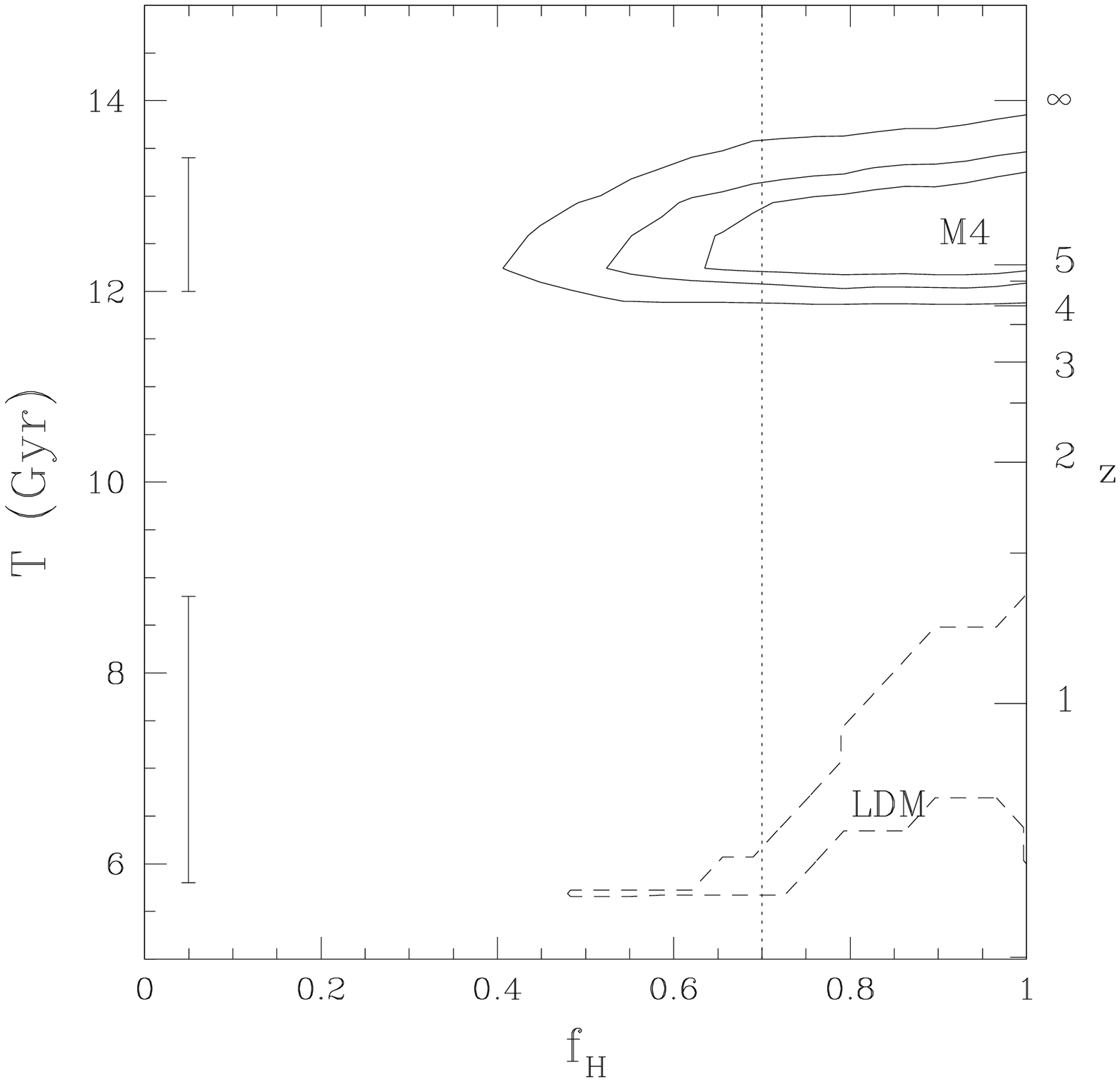}
\clearpage
\plotone{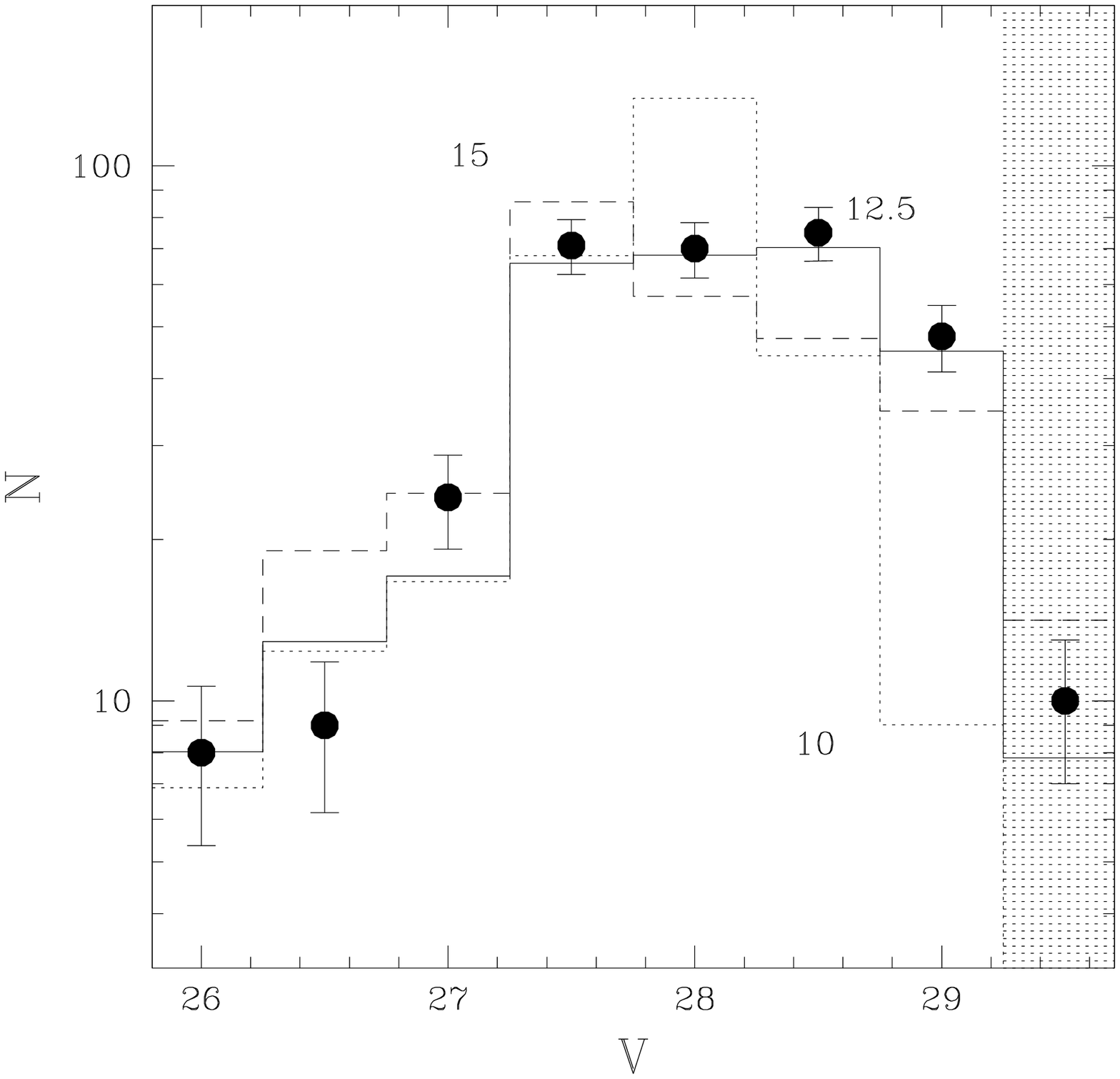}

\begin{references}
\reference{BRL} Bergeron, P., Ruiz, M. T. \& Leggett, S. K. 1998, ApJ, 108, 339
\reference{BDB} Binney, J., Dehnen, W. \& Bertelli, G., 2000, MNRAS, 318, 658
\reference{BH} Bolte, M. C. \& Hogan, C. J., 1995, Nature, 376, 399
\reference{BTH} Burkert, A., Truran, J. W. \& Hensler, G., 1992, ApJ, 391, 651
\reference{CGC} Carraro, G., Girardi, L. \& Chiosi, C., 1999, MNRAS, 309, 430
\reference{CGC} Carretta, E., Gratton, R. G., Clementini, G. \& Fusi Pecci, F., 2000,
ApJ, 533, 215
\reference{CCd} Cassisi, S., Castellani, V., degl'Innocenti, S., Salaris, M. \&
Weiss, A., 1999, A\&AS, 134, 103
\reference{CH} Cayrel, R.,  {\it et al.} , 2001, Nature, 409, 691
\reference{C01} Chaboyer, B. C., 2001, in ASP Conf. Ser. 245, Astrophysical Ages and Time scales,
ed. von Hippel, T., Simpson, C. \& Manset, N., (San Francisco: ASP), 307
\reference{CDKK} Chaboyer, B. C., Demarque, P., Kernan, P. J. \& Krauss, L. M., 1998, ApJ, 494, 96
\reference{EGSC} Elson, R. A., Gilmore, G. F., Santiago, B. X. \& Casertano, S., AJ, 110, 682
\reference{FBB} Frye, B., Broadhurst, T. \& Benitez, N., 2001, /astro-ph/0112095
\reference{JFK} Jimenez, R., Flynn, C. \& Kotonova, 1998, MNRAS, 299, 515
\reference{G86} Greenstein, J., 1986, ApJ, 304, 334
\reference{H99} Hansen, B. M. S., 1999, ApJ, 520, 680
\reference{H01} Hansen, B., M. S., 2001, ApJ, 558, L39
\reference{HPT} Hurley, J. R., Pols, O. R.\& Tout, C. A., 2000, MNRAS, 315, 543
\reference{KHH} Knox, R. A., Hawkins, M. R. S. \& Hambly, N. C., 1999, MNRAS, 306, 736
\reference{KB} Koopmans, L. \& Blandford, R. D., 2001, /astro-ph/0107358
\reference{LRB} Leggett, S. K., Ruiz, M. T. \& Bergeron, P., 1998, ApJ, 497, 294
\reference{LDM} Liebert, J. Dahn, C. C. \& Monet, D. G., 1988, ApJ, 332, 891
\reference{LC} Liu, W. M. \& Chaboyer, B., 2000, ApJ, 544, 818
\reference{MFL} Monet, D. G., Fisher, M. D., Liebert, J., Canzian, B., Harris, H. C., \& Reid, I. N., 2000, AJ 120, 1541
\reference{NB} Ng, Y. K., \& Bertelli, G., 1998, A\&A, 329, 943
\reference{OHD} Oppenheimer, B. R., Hambly, N. C., Digby, A. P., Hodgkin, S. T. \& Saumon, D., 2001,
Science, 292, 698
\reference{OR} Ortolani \& Rosino, L. 1987, A\&A, 185, 102
\reference{Os} Oswalt, T. D., Smith, J. A., Wood, M. A. \& Hintzen, P. M., 1996, Nature, 382, 692
\reference{PDR} Paresce, F., De Marchi, G. \& Romaniello, M., ApJ, 440, 216
\reference{PW} Prochaska, J. X. \& Wolfe, A. M., 1997, ApJ, 487, 73
\reference{RSH} Reid, I. N., Sahu, K. C. \& Hawley, S. L., 2001, ApJ, 559, 942
\reference{R78} Richer, H. B., 1978, ApJ, 224, L9
\reference{RF} Richer, H. B. \& Fahlman, G. G., 1988, ApJ, 325, 218
\reference{RFI} Richer, H. B.,  {\it et al.} , 1995, ApJ, 451, L17
\reference{R97} Richer, H. B.,  {\it et al.} , 1997, ApJ, 484, 741
\reference{R02} Richer, H. B.,  {\it et al.} , 2002, {\bf the MS paper}
\reference{Sand} Sandage, A., 1953, AJ, 58, 61
\reference{SCD} Sarajedini, A., Chaboyer, B. \& Demarque, P., 1997, PASP, 109, 1321
\reference{SC} Segretain, L. \& Chabrier, G., 1993, A\&A, 271, L13
\reference{SSA} Shapley, A. E., Steidel, C. C., Adelberger, K. L., Dickinson, M., Giovalisco, M.
\& Pettini, M., 2001, ApJ, 562, 95
\reference{S84} Sion, E. M., 1984, ApJ, 282, 612
\reference{SVB} Stetson, P. B., Vandenberg, S. \& Bolte, M., 1996, PASP, 108, 560
\reference{VDS} Van Dokkum, P. G. \& Stanford, S. A., 2001, ApJ, 562, L35
\reference{Vo} Vogt, N. P.,  {\it et al.} , 1997, ApJ, 479, L121
\reference{Win} Winget, D.  {\it et al.} , 1987, ApJ, 315, L77
\reference{WP} Wolfe, A. M. \& Prochaska, J. X., 2000, ApJ, 545, 591
\reference{Wood} Wood, M. A., 1992, ApJ, 386, 539
\end{references}
\end{document}